\documentclass[journal=jctcce,manuscript=article,etalmode=truncate,maxauthors=4]{achemso}
\setkeys{acs}{articletitle = true}

\usepackage[utf8]{inputenc}
\usepackage[T1]{fontenc}
\usepackage{mathptmx}
\usepackage{graphicx}
\usepackage[usenames,dvipsnames]{xcolor}
\usepackage{siunitx}
\usepackage{enumitem}
\usepackage{amsmath}
\usepackage{amssymb}
\usepackage{dsfont}
\usepackage{tabularx}
\usepackage{tikz}
\usepackage{chemfig}
\usepackage{stackengine}
\usepackage{booktabs}
\usepackage{threeparttable}
\usepackage[caption=false]{subfig}
\usepackage{hyperref}

\DeclareMathAlphabet{\mathcal}{OMS}{cmsy}{m}{n}

\SectionNumbersOn
\title{On the equivalence of the hybrid particle\textendash field and Gaussian core models}

\author{Morten Ledum}  
\affiliation{
    Department of Chemistry, and Hylleraas Centre for Quantum Molecular Sciences, University of Oslo, PO Box 1033 Blindern, 0315 Oslo, Norway
} 
  
\author{Samiran Sen} 
\affiliation{
    Department of Chemistry, and Hylleraas Centre for Quantum Molecular Sciences, University of Oslo, PO Box 1033 Blindern, 0315 Oslo, Norway
}

\author{Sigbj\o rn L\o land Bore}
\affiliation{
    Department of Chemistry, and Hylleraas Centre for Quantum Molecular Sciences, University of Oslo, PO Box 1033 Blindern, 0315 Oslo, Norway
}
  
\author{Michele Cascella} 
\email{michele.cascella@kjemi.uio.no}
\affiliation{
    Department of Chemistry, and Hylleraas Centre for Quantum Molecular Sciences, University of Oslo, PO Box 1033 Blindern, 0315 Oslo, Norway
}

\begin{document}
\maketitle
\clearpage 

\begin{abstract}
    Hybrid particle\textendash field molecular dynamics is a molecular simulation strategy wherein particles couple to a density field instead of through ordinary pair potentials. Traditionally considered a mean-field theory, a momentum and energy-conserving hybrid particle\textendash field formalism has recently been introduced, which was demonstrated to approach the Gaussian Core model potential in the grid-converged limit. Here, we expand on and generalize the correspondence between the Hamiltonian hybrid particle\textendash field method and particle\textendash particle pair potentials. Using the spectral procedure suggested by Bore and Cascella, we establish compatibility to \textit{any} local soft pair potential in the limit of infinitesimal grid spacing. Furthermore, we document how the mean-field regime often observed in hybrid particle\textendash field simulations is due to the systems under consideration, and not an inherent property of the model. Considering the Gaussian filter form in particular, we demonstrate the ability of the Hamiltonian hybrid particle\textendash field model of recovering all structural and dynamical properties of the Gaussian Core model, including solid phases, a first-order phase transition, and anomalous transport properties. We quantify the impact of the grid spacing on the correspondence, as well as the effect of the particle\textendash field filtering length scale on the emergent particle\textendash particle correlations.
\end{abstract}
\newpage 

\section{Introduction}
Hybrid particle\textendash field (hPF) methods are a group of schemes for efficient molecular resolution simulations of soft matter systems. In hPF, the computationally expensive non-bonded pair interactions of traditional molecular dynamics approaches are done away with, instead coupling particles not explicitly bonded to each other only through interactions with a density field. In the last two decades, these strategies have been employed to simulate a wide range of soft matter systems through the Monte Carlo-based \textit{single chain in mean-field} (SCMF) method developed by M\"uller and co-workers\cite{daoulas2006single,daoulas2006morphology} and by molecular dynamics (MD).\cite{Milano2009JCP,milano2013hybrid} Through coarse particle representations and quasi-instantaneous approximations of the time-evolution of the density field, hPF approaches have been successful in modelling many important phenomena, including lamellar and non-lamellar lipid phases,\cite{de2013improved,ledum2020automated,Denicola2020BBA,Soares2017JPCL,C6SM02252A,C7CP03871B} polyelectrolytes,\cite{Zhu2016PCCP,kolli2018JCTC,Bore2019JCTC} polypeptides,\cite{bore2018hybrid} nanoparticles,\cite{Munao2018NANOSCALE} carbon nanotubes\cite{Zhao2016NANOSCALE}, and self-assembly of block copolymers~\cite{SCHNEIDER2019463}.

The original derivation of the hPF scheme relied upon a self-consistent field theoretical approach and the mean-field approximation through the application of a saddle point approximation. 
In fact, to what extent the mean-field approximations introduced in the formal derivation of the hPF-MD  equations affect the sampled conformational space, and more directly, MD equations, have not been rigorously discussed in the past literature.
In general, it has been accepted that, in an hPF simulation, particles move rather independently, producing weakly structured liquids.

Traditionally in hPF approaches, a particle\textendash mesh strategy is employed wherein molecules are distributed on a coarse computational grid according to a distribution window function, e.g.\ \textit{cloud-in-cell}\cite{hockney2021computer} (CIC). Subsequently, the force density is calculated as finite difference derivatives of the density grid at each vertex using a staggered grid setup.\cite{Zhao2012JCP} This naturally lends itself to efficient parallelization because of the time scale dichotomy; the coarse density grid is very slowly varying compared to the individual microscopic particle positions. This approach necessarily entangles the spacing of the grid to the level of coarsening of the model, thus hPF calculations must compromise between the wanted density spread and the numerical accuracy of the grid operations. Particle\textendash particle correlations have been shown to arise in canonical hPF-MD simulations with very short grid spacing, displaying some limited ability to recover radial distribution functions of coarse-grained (CG) MD simulations\cite{Milano2009JCP,Nicolia2011JCTC} However, such grid setups are rarely if ever used in production runs, as using grids with a sufficient number of vertices for this to present is prohibitively expensive, and their usage curtails the intrinsic strength of the hPF approach\textemdash its parallelizability and speed. Additionally, it may create numerical stability issues related to excessively fast oscillations in the density function. As such, the majority of hPF-MD results reported in the literature lie solidly in what appears as a \textit{mean-field} regime where steric interactions are poorly represented. The literature also lacks a proper discussion on the ability of hPF models to respect mechanical conservation laws, and whether any discrepancy might be connected to such approximation, either on a formal level or through the algorithmic implementation. 

Recently, stemming from seeding work by Theodorou and coworkers,\cite{Vogiatzis2017MACRO} two of us presented a reformulation of the hPF scheme\cite{bore2020hamiltonian}  foregoing entirely the statistical mechanics arguments, obtaining instead  the forces acting on the particle directly from spatial derivatives of the interaction density functional. The new formalism defines a direct connection between the individual microscopic states of the system and its energetics, in full accord with standard Hamiltonian mechanics. In the following, we will refer to this new framework as the Hamiltonian-hPF (HhPF).


HhPF profits from the employment of a grid-independent filtering function $\mathcal{G}(\mathbf{x})$ that determines the intrinsic density spread associated with each particle, separately from the window function $P$, which assigns particles to the computational grid,
\begin{equation}
    \tilde\phi(\mathbf{r})\equiv \int\mathrm{d}\mathbf{x}\, \phi(\mathbf{x})\mathcal{G}(\mathbf{r}-\mathbf{x}), \quad\mathrm{with}\quad \phi(\mathbf{r}) = \sum^N_{i=1} P(\mathbf{r}-\mathbf{r}_i).
\end{equation}
By decoupling the definition of the density spread from the computational grid, systematic convergence of the numerical force calculations is possible. 
Moreover, by propagating in parallel the fast intramolecular and the slow density-field forces via a multiple time-step algorithm, we showed that it is possible to retain rigorous, well-behaved dynamics with strict conservation of the energy.\cite{ledum2021hylleraasmd} The HhPF framework represents a type of simplified long-range Ewald approach. Such particle\textendash mesh methods have a long history in atomic simulations,\cite{hockney2021computer} most notably through the ubiquitous particle\textendash mesh Ewald (PME) approach to electrostatic interactions.\cite{Darden1993Particle} Beyond this, a large body of work dating back from the early 1970s involves employing Ewald summation techniques to compute the long-range interactions of the Lennard-Jones (L-J) potential.\cite{karasawa1989acceleration,shi2006molecular} The $r^{-6}$ term is usually handled by cut-off strategies, but in some cases, the long-range dispersion part of the L-J potential is particularly important and should not be neglected, e.g.\ for interfacial phenomena.\cite{in2007application}

Thanks to its well-behaved numerical convergence, HhPF allows for a systematic exploration of all mechanical properties of hPF models, and in particular, their relationship to standard particle-particle models interacting by two-body potentials. In fact, for a pertinent choice of the filtering function,\cite{laradji1994off,pike2009theoretically,bore2020hamiltonian} we argued that it was possible to establish a direct link between the HhPF model and the Gaussian Core model (GCM) of purely repulsive interparticle interactions by Stillinger.\cite{stillinger1976phase} 

In this work, we extend and generalize the preliminary result reported in Ref.~\cite{bore2020hamiltonian} showing that grid-converged filtered spectral hPF models fully describe particle\textendash particle correlation, and in particular, that they can recover the behaviour of \textit{any}  soft-core pair potential. As a natural example, we consider, for careful study, the GCM model, showing how the HhPF reproduces exactly the GCM's microscopic dynamics. Moreover, we show how the strength of particle\textendash particle correlations is controlled by the HhPF filtering length scale, and that the apparent mean-field regime observed in hPF simulations must be understood as an approximation of the behaviour of an interacting particle model in the high temperature / weak interactions limit.

\section{Theory and methods}
\subsection{Hamiltonian hybrid particle-field}
In the following, we briefly present the HhPF method. For a thorough derivation of the hPF and HhPF schemes, the reader is directed to Refs.~\citenum{Milano2009JCP} and \citenum{bore2020hamiltonian}. We consider a system of $N$ particles in $M$ molecules subject to the Hamiltonian
\begin{equation}\label{eq:H}
    \mathcal{H}(\mathbf{R}, \mathbf{P})=\sum_{m=1}^M \mathcal{H}_0(\mathbf{R}^m, \mathbf{P}^m) + W[\tilde{\phi}(\mathbf{r})],
\end{equation}
where particle positions $\mathbf{R}=\{\mathbf{r}_i\}_{i=1}^N$ and momenta $\mathbf{P}=\{\mathbf{p}_i\}_{i=1}^N$ fully specify microstates of the system, and $\mathbf{R}^m$ and $\mathbf{P}^m$ indicate the positions and momenta associated with particles in molecule $m$. Here, $\mathcal{H}_0$ is the Hamiltonian of a single non-interacting molecule $m$, and $W$ is an interaction energy functional depending on the filtered particle number densities $\tilde\phi(\mathbf{r})$,
\begin{equation}\label{eq:phi}
    \tilde\phi(\mathbf{r})\equiv \int\mathrm{d}\mathbf{x}\, \phi(\mathbf{x})\mathcal{G}(\mathbf{r}-\mathbf{x}), \quad \phi(\mathbf{r}) = \sum^N_{i=1} P(\mathbf{r}-\mathbf{r}_i),
\end{equation}
where $\mathcal{G}$ is a filtering function, and $P$ is a window function used to distribute the particles in the space. 

The sampling of the phase space associated with Eq.\ \ref{eq:H} using MD requires computing the forces due to both $\mathcal{H}_0$ and $W$. The forces due to bonded interactions terms of single molecules, $\mathcal{H}_0(\mathbf{R}^m)$, are computed by
\begin{equation}
    \mathbf{F}_{i}^\mathrm{bonded}=-\frac{\partial \mathcal{H}_0(\mathbf{R}^m)}{\partial \mathbf{r}_i}.
\end{equation}
Forces due to particle\textendash field interactions are obtained by
\begin{equation}\label{eq:hpf-force}
\mathbf{F}_{i}^\text{HhPF} = - \int\mathrm{d}\mathbf{r}\, \nabla V(\mathbf{r}) P(\mathbf{r}-\mathbf{r}_i), \quad V(\mathbf{r}) = \int\mathrm{d}\mathbf{y}\,\frac{\partial w}{\partial \tilde\phi}(\mathbf{y})\mathcal{G}(\mathbf{r}-\mathbf{y}).
\end{equation}
Here, $V$ is the external potential acting on the particles, and lower-case $w$ denotes the interaction energy density, with $W = \int w(\tilde{\phi}({\mathbf{r}})) \text d{\mathbf{r}}$.

In practice, the estimation of discrete densities is done using a CIC window function $P$, which distributes particles on the nearest grid points by trilinear interpolation. The density is computed at the discretized $(n, m, \ell)$ grid point at position $\mathbf{r}_{nm\ell}$ by
\begin{equation}
    \phi_{nm\ell} \equiv \phi(\mathbf{r}_{nm\ell}) = \sum^N_{i=1}P(\mathbf{r}_{nm\ell}-\mathbf{r}_i).
\end{equation}

The spectral method suggested by Bore and Cascella\cite{bore2020hamiltonian} dictates performing the convolution of equation \ref{eq:phi} in reciprocal space, via
\begin{equation}
    \tilde \phi=\text{FFT}^{-1}\left[\text{FFT}(\phi)\text{FFT}(\mathcal{G})\right], \label{eq:phi-fft}
\end{equation}
with the external potential computed as:
\begin{equation}
    V = \text{FFT}^{-1}\left[\text{FFT}\left(\frac{\partial w(\tilde\phi(\mathbf{r}))}{\partial \tilde\phi}\right)\text{FFT}(\mathcal{G})\right]. \label{eq:v-fft}
\end{equation}
The derivative of $V$ is further computed in reciprocal space, 
\begin{equation}
    \nabla V =\text{FFT}^{-1}\left[i\mathbf{k}~\text{FFT}\left(\frac{\partial w(\tilde\phi(\mathbf{r}))}{\partial \tilde\phi}\right)\text{FFT}(\mathcal{G})\right]. \label{eq:nabla-v-fft}
\end{equation}
Note that grid-indices have been suppressed in Eqs.\ \ref{eq:phi-fft}, \ref{eq:v-fft}, and \ref{eq:nabla-v-fft}, but these operations are performed on the full computational grid. From the discretized $\nabla V_{nm\ell}\equiv \nabla V(\mathbf{r}_{nm\ell})$, the forces are computed by interpolating back onto particle positions $\mathbf{r}_i$ through equation \ref{eq:hpf-force} by
\begin{equation}\label{eq:F-P4}
    \mathbf{F}_i^\mathrm{HhPF} = -\sum_{nm\ell} \nabla V_{nm\ell} P(\mathbf{r}_{nm\ell}-\mathbf{r}_i)h^3,
\end{equation}
where $h^3$ is the volume of a single cell. The interpolating sum is taken over all grid points, however the window function $P$ centered at $\mathbf{r}$ normally vanishes on any grid vertex which is not in the immediate vicinity of $\mathbf{r}$.

\subsection{Gaussian core model}
The Gaussian core model (GCM), introduced by Stillinger et al.\cite{stillinger1976phase}, defines a purely repulsive particle\textendash particle interaction in the functional shape of a Gaussian,
\begin{equation}
    V(r)=\varepsilon\times\exp\left(\frac{-r^2}{\sigma^2}\right),
\end{equation}
where $r\equiv \vert \mathbf{r}_i-\mathbf{r}_j\vert$ denotes the $i$\textendash $j$ inter-particle distance, $\varepsilon$ is an energy scale, and $\sigma$ defines the effective interaction length. As we will make clear shortly, the Gaussian core interaction can alternatively be viewed as the overlap integral between the densities of two smeared particles of Gaussian spreads (with variance $\sigma^2/2$), with centers separated by a distance $r$.

In terms of the reduced density $\rho^*=\rho\sigma^3$ and reduced temperature $T^*=Tk_\mathrm{B}/\varepsilon$, the GCM admits thermodynamically stable solid phases in the low-temperature regime, $T^*<0.01$.\cite{prestipino2005phase,lang2000fluid} In fact, despite its deceptively simple form, the GCM exhibits a range of interesting and anomalous dynamical and structural properties, including increasing isothermal diffusivity and shear viscosity for increasing density,\cite{mausbach2009transport} reentrant compression melting at constant temperature and a maximum freezing temperature,\cite{stillinger1978study} hard-sphere behaviour in the limit of vanishing temperature and density,\cite{stillinger1976phase} and negative thermal expansion.\cite{stillinger1997negative}

\subsection{Simulation details and reduced units} 
Here we outline all the details necessary to reproduce the figures and results contained in this work. Unless otherwise noted, all results presented have scaled lengths, energies, and masses scaled to unity, $k_\text{B}=\sigma=\varepsilon=m=1$, with reduced temperature $T^*$ and reduced density $\rho^*$ uniquely determining the chemical state. Times are measured in the characteristic time scale of the GCM, $t^*=\sqrt{m\sigma^2/\varepsilon}$. Apart from the one-dimensional test cases, all HhPF simulations are run using HylleraasMD\cite{ledum2021hylleraasmd}. The corresponding GCM simulations are performed in LAMMPS\cite{LAMMPS}.

The one-dimensional simulations 
are calculated with HhPF using an idealized system of two particles separated by a distance $r$ on a computational grid of $2^{11}$ grid points across a simulation box length $L\gg\sigma$, with a real space grid spacing of $0.0073$. 

For the force comparison, 
a system of $N$ particles varying from $800$ to $5250$ with corresponding reduced densities between $\rho^*=0.1$ and $\rho^*=0.66$ is used with the force deviation being averaged over all particles in the simulation. The forces are calculated from a single snapshot of a random particle conformation in each case for each value of the grid spacing $h$.

The solid/liquid systems 
are constituted by 1458 (1372) particles for the $\rho^*=0.25$ and $\rho^*=1.0$ ($\rho^*=0.12$) cases with grid spacings $h=0.2813$ and $h=0.1771$ ($h=0.352$). In each case, the temperature is held at $T^*=0.001$ using a CSVR thermostat\cite{Bussi2007JCP} with coupling strength $\tau=0.1$ for $n=5\cdot10^4$ MD steps of time step $\Delta t=0.01$. The $\rho^*=0.25$ BCC crystal system presented exhibiting the first-order phase transition is the same as that of the previously mentioned $\rho^*=0.25$ system, being slowly cooled using a CSVR thermostat ($\tau=0.1$) from $T^*=0.02$ to $T^*=0.006$ over $n=10^6$ time steps of $\Delta t=0.025$ with grid spacing $h=0.0703$. The GCM critical temperature and latent heat of melting is estimated from a corresponding GCM simulation of the same reduced density and the same number of particles.

The dependency of the radial distribution function on the numerical grid spacing was monitored on 
the $\rho^*=0.12$ system at a temperature $T^*=0.001$, running $n=5\cdot10^4$ MD steps with $\Delta t=0.05$ at varying grid spacings.

The velocity autocorrelations were estimated from simulations of systems containing $N=1041$ particles, at number density $\rho=8.33$, with $T^*=0.008$, and varying values of the HhPF filtering widths, between  $\sigma_\mathrm{hPF}=0.1$ and $\sigma_\mathrm{hPF}=1.0$. The diffusion constants were calculated from $N=1458$ particle simulations with reduced densities varying between $\rho^*=0.1$ to $\rho^*=1.0$.


\section{Results and discussion}
\subsection{Equivalence of the Hamiltonian hybrid particle\textendash field method and any finite pair potential}
As evidenced by Bore and Cascella,\cite{bore2020hamiltonian} a quadratic density dependence in $W[\tilde\phi]$, e.g.\ the commonly used interaction term
\begin{equation}
    W[\tilde\phi] = \frac{1}{2\phi_0}\int\mathrm{d}\mathbf{r}\, \tilde\chi_{AB}\tilde\phi_A(\mathbf{r})\tilde\phi_B(\mathbf{r}), \label{eq:Wchi}
\end{equation}
allows establishing a correspondence between any potential form and the hPF formalism through the filtering function. Here, $\tilde\chi_{AB}$ is the interaction energy parameter between species $A$ and $B$, and $\phi_0$ is the average number density across the simulation domain.

In the following, let $K(\mathbf{x}-\mathbf{y})$ denote any local, almost everywhere finite, absolutely-, and square-integrable function over the simulation volume. Let the interaction be modulated by an energy scale $\varepsilon$,
\begin{equation}
    W_K[\tilde\phi] 
    = \frac{\varepsilon}{2\phi_0}\iint\mathrm{d}\mathbf{r}\mathrm{d}\mathbf{r}'\, \tilde\phi_A(\mathbf{r})K(\mathbf{r}-\mathbf{r}')\tilde\phi_B(\mathbf{r}') 
    = \frac{\varepsilon}{2\phi_0}\int\mathrm{d}\mathbf{r}'\,\left[\int\mathrm{d}\mathbf{r}\, \tilde\phi_A(\mathbf{r})K(\mathbf{r}-\mathbf{r}')\right]\tilde\phi_B(\mathbf{r}').
\end{equation}
With $\mathcal{F}$ denoting the Fourier transform, applying the convolution theorem to the inner integral yields
\begin{equation}
    W_K[\tilde\phi] 
    = \frac{\varepsilon}{2\phi_0}\int\mathrm{d}\mathbf{r}'\,\tilde\phi_B(\mathbf{r}')\mathcal{F}^{-1}\left[\tilde\phi_A(\mathbf{k})K(\mathbf{k})\right].
\end{equation}
Noting that in periodic boundary conditions, any filtered density is necessarily an even function and the $r=\vert \mathbf{r}-\mathbf{r}'\vert$-dependent pair potential $K(\mathbf{r}-\mathbf{r}')$ is real and symmetric, the Fourier transformed convolution is guaranteed to be real. Application of the Plancherel theorem thus gives
\begin{equation}
    W_K[\tilde\phi] 
    = \frac{\varepsilon}{2\phi_0}\int\mathrm{d}\mathbf{k}\,\tilde\phi_A(\mathbf{k})K(\mathbf{k})\tilde\phi_B(\mathbf{k}). \label{eq:KH2equivalent}
\end{equation}
Recall that $\tilde\phi(\mathbf{r})=\int\mathrm{d}\mathbf{r'}\,\mathcal{G}(\mathbf{r}-\mathbf{r}')\phi(\mathbf{r'})$. Together with equation \ref{eq:Wchi}, this establishes a correspondence between the squared filter in reciprocal space, $\mathcal{G}(\mathbf{k})^2$, and the Fourier transformed pair potential, $K(\mathbf{k})$. 

\subsection{Real-space derivation of the hybrid particle\textendash field\textendash Gaussian core model equivalency}
Evidently, it is possible to identify a filtering function\textendash pair potential link in reciprocal space for any suitable soft-core potential $K(\mathbf{r}-\mathbf{r}')$. However, applying a Fourier space square root in no way guarantees that the resulting filter function is easily representable in direct space as a closed-form expression, and neither does it guarantee that the resulting square root is real. One notable exception, however, is the Gaussian functional form, which survives both Fourier transforms and exponentiation (in real-, or Fourier space) without changing its fundamental character. 

In the following, assume the filtering length is much smaller than the simulation box size, $\sigma\ll L$. In this regime, we extend real-space interaction integrals to cover $\mathds{R}^3$, and ignore self-interaction between particles and periodic images of the same particle. Under this assumption, the Gaussian-filtered hPF interaction energy integral
\begin{equation}
    W[\tilde\phi] = W_\chi[\tilde\phi] + W_\kappa[\tilde\phi] = \frac{1}{\phi_0}\int\mathrm{d}\mathbf{r}\, 
    \sum_{i<j}\tilde{\chi}_{ij}\tilde{\phi}_i(\mathbf{r})\tilde{\phi}_j(\mathbf{r})
    +
    \frac{1}{2\kappa \phi_0}\int\mathrm{d}\mathbf{r}\, 
    \left(\sum_{i}\tilde{\phi}_i(\mathbf{r})-\phi_0\right)^2 \label{eq:hamiltonian}
\end{equation}
becomes analytically tractable in direct space. 

The microscopic number density arising from $N$ particles of type $k$ at positions $\{\mathbf{r}_\ell\}_{\ell=1}^N$ is given by
\begin{equation}
    \phi_k(\mathbf{r})=\sum_{\ell=1}^N\delta(\mathbf{r}-\mathbf{r}_\ell),
\end{equation}
with the \textit{filtered} density 
\begin{equation}
    \tilde{\phi}_k(\mathbf{r})=\sum_{\ell=1}^N\frac{1}{\sqrt{8\pi^3}\sigma^3}\exp\left[\frac{-(\mathbf{r}-\mathbf{r}_\ell)^2}{2\sigma_\mathrm{hPF}^2}\right].
\end{equation}
Using the \textit{Gaussian product rule},\cite{boys1950electronic} the $\tilde{\phi}_i(\mathbf{r})\tilde{\phi}_j(\mathbf{r})$ product in the $W_\chi$-term
may be expressed as
\begin{equation}
    \tilde{\phi}_i(\mathbf{r})\tilde{\phi}_j(\mathbf{r})=\frac{1}{8\pi^3\sigma_\mathrm{hPF}^6}\sum_{k=1}^M\sum_{\ell=1}^N\exp\left[\frac{-(\mathbf{r}_k-\mathbf{r}_\ell)^2}{4\sigma_\mathrm{hPF}^2}\right] \exp\left[\frac{-\left(\mathbf{r}-\left[\frac{\mathbf{r}_k+\mathbf{r}_\ell}{2}\right]\right)^2}{\sigma_\mathrm{hPF}^2}\right].
\end{equation}
Integrating such terms yields the $W_\chi$-term of the interaction energy integral as
\begin{equation}
\begin{aligned}
    \int\mathrm{d}\mathbf{r}\,\tilde{\phi}_i(\mathbf{r})\tilde{\phi}_j(\mathbf{r})&=\frac{1}{8\pi^3\sigma_\mathrm{hPF}^6}\sum_{k=1}^M\sum_{\ell=1}^N\exp\left[\frac{-(\mathbf{r}_k-\mathbf{r}_\ell)^2}{4\sigma_\mathrm{hPF}^2}\right] \int\mathrm{d}\mathbf{r}\,\exp\left[\frac{-\left(\mathbf{r}-\left[\frac{\mathbf{r}_k+\mathbf{r}_\ell}{2}\right]\right)^2}{\sigma_\mathrm{hPF}^2}\right] \\
    &=\frac{1}{8\sqrt{\pi^3}\sigma_\mathrm{hPF}^3}\sum_{k=1}^M\sum_{\ell=1}^N\exp\left[\frac{-(\mathbf{r}_k-\mathbf{r}_\ell)^2}{4\sigma_\mathrm{hPF}^2}\right]. \label{eq:intphiphi}
\end{aligned}
\end{equation}

The $W_\kappa$-term contains three contributions: The constant contributes a factor proportional to the simulation box volume, while the cross-term contains integrals of the form $\int\mathrm{d}\mathbf{r}\,\tilde{\phi}_k(\mathbf{r})$ which trivially evaluate to unity by design of the normalization. The $\int\mathrm{d}\mathbf{r}\,\tilde{\phi}_k(\mathbf{r})\tilde{\phi}_k(\mathbf{r})$ term contributes a term of the same form as $W_\chi$.

In full, the interaction energy is given by 
\begin{equation}
\begin{aligned}
    W[\tilde\phi]&=\frac{1}{8\sqrt{\pi^3}\sigma_\mathrm{hPF}^3\phi_0}\sum_{i=1}^K\sum_{j=i+1}^K\sum_{k=1}^{N_i}\sum_{\ell=1}^{N_j}\tilde\chi_{ij}\exp\left[\frac{-(\mathbf{r}_k-\mathbf{r}_\ell)^2}{4\sigma_\mathrm{hPF}^2}\right],\\
    &\qquad\qquad +\frac{1}{16\sqrt{\pi^3}\sigma_\mathrm{hPF}^3\kappa\phi_0}\sum_{i=1}^K\sum_{k=1}^{N_k}\sum_{\ell=1}^{N_k}\exp\left[\frac{-(\mathbf{r}_k-\mathbf{r}_\ell)^2}{4\sigma_\mathrm{hPF}^2}\right] \\
    &\qquad\qquad - \frac{N}{2\kappa},
\end{aligned}
\end{equation}
where $K$ is the number of different moieties, each with $N_k$ particles. In summary, apart from a constant energy shift that does not impact dynamics, the Gaussian-filtered HhPF corresponds to a GCM with $\sigma^2=4\sigma_\mathrm{hPF}^2$, and 
\begin{align}
    \varepsilon=\frac{1}{16\sqrt{\pi^3}\sigma_\mathrm{hPF}^3\phi_0}\left(2\tilde\chi_{ij}+\frac{1}{\kappa}\right).
\end{align}


\subsection{Grid-converged HhPF recovers \textit{any} soft-core pair potential}
\begin{figure}[t]
    \centering
    \includegraphics[width=0.40\textwidth]{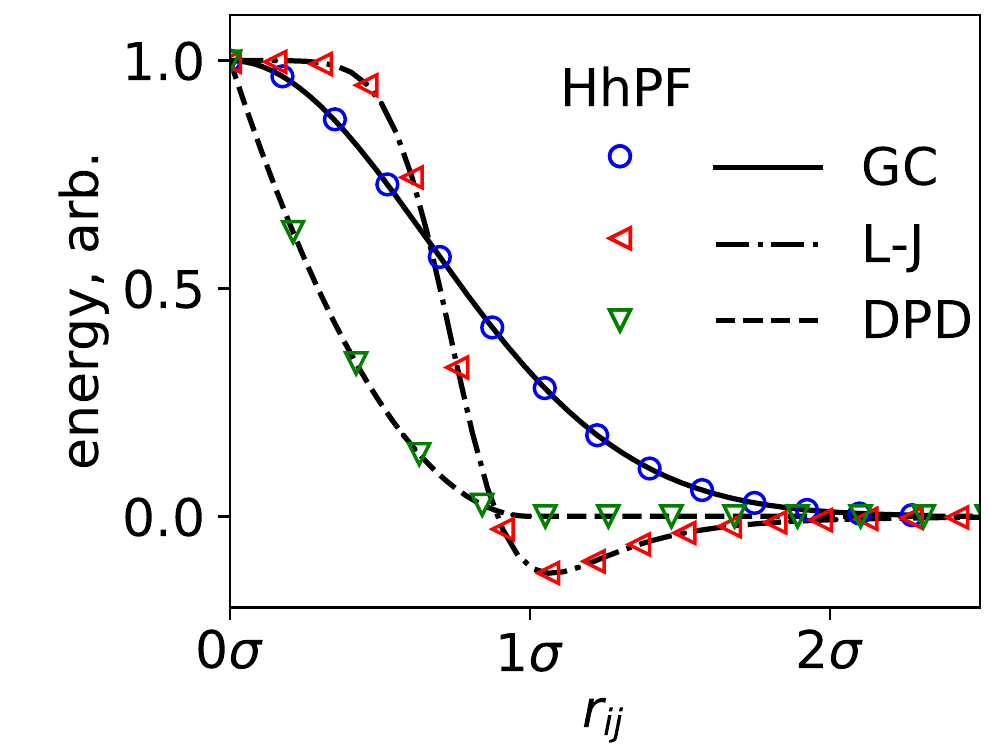}
    \includegraphics[width=0.40\textwidth]{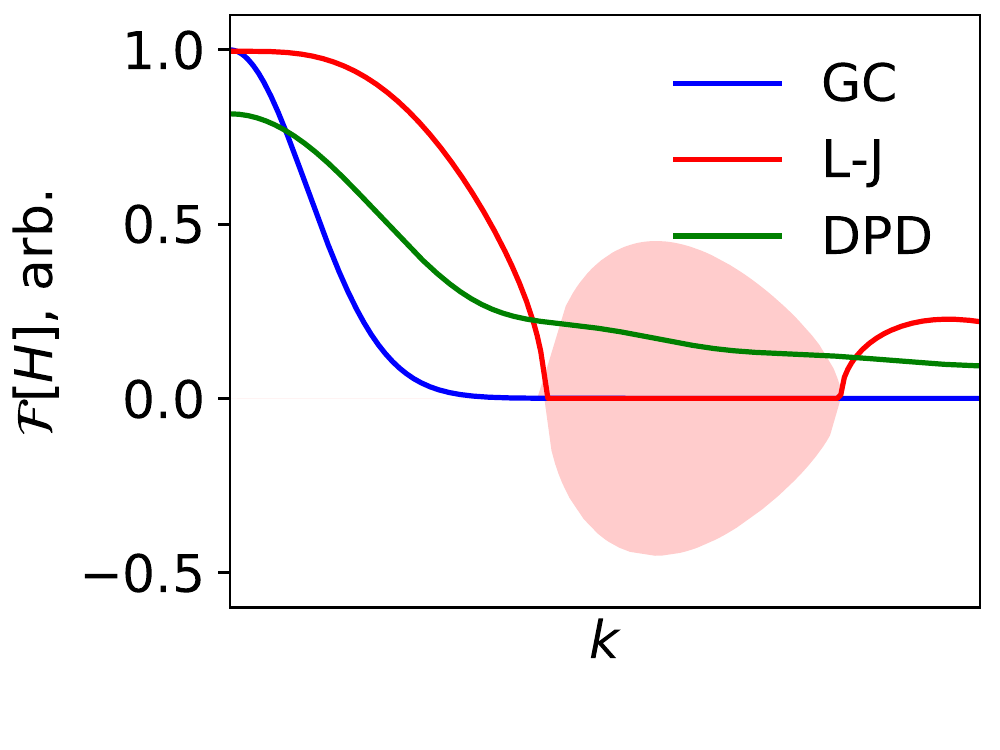}
    \caption{Example correspondence between filtered hPF interactions arising from using the square root of $\mathcal{F}[V(\mathbf{r}-\mathbf{r}')]$ as the filter. Left: Energies for a one-dimensional example HhPF in the small grid length limit, comparison with explicit particle\textendash particle potentials. Right: Reciprocal space representation of the filters used. The shaded red area represents the imaginary part of the soft-core Lennard-Jones filter, containing a rapidly oscillating region.}
    \label{fig:potentials}
\end{figure}

The correspondence between the filtered hPF approach and \textit{any} soft local pair potential $K$ is demonstrated explicitly in Figure~\ref{fig:potentials}, where a few simple one-dimensional examples are shown. The dissipative particle dynamics (DPD) potential is 
\begin{equation}
    V_\mathrm{DPD}(r_{ij})=\left(1-\frac{r_{ij}}{\sigma}\right)^2,
\end{equation}
commonly used also as the $\omega^R(r_{ij})$ potential which modulates the random DPD Wiener process. The soft-core version of the Lennard-Jones potential used here takes the form 
\begin{equation}
    V_{\mathrm{Soft}\ \mathrm{L-J}}(r_{ij})=\frac{\sigma^{12}}{(r_{ij}^6+\frac{1}{2})^2}-\frac{\sigma^6}{r_{ij}^6+\frac{1}{2}}.
\end{equation}
In both cases, $\sigma$ denotes an arbitrary length scale. On the right-hand side of Figure~\ref{fig:potentials} is shown the reciprocal space representation of the filters used, resulting from the square root of the Fourier transformed pair potential, $H^2(k)=K(k)$. Evidently, the potential energy of the explicit pair-potentials is matched to arbitrary precision by the HhPF scheme, provided the corresponding filtering function is used.

We note that in the case of the soft-core Lennard-Jones potential, the rapidly oscillatory region (the negative, attractive part) introduces instabilities in the HhPF model at very small ($h\lesssim0.005$) grid spacings.

\subsection{Dynamical properties of HhPF and GCM} 
\begin{figure}[t]
    \centering
        \includegraphics[width=0.40\textwidth]{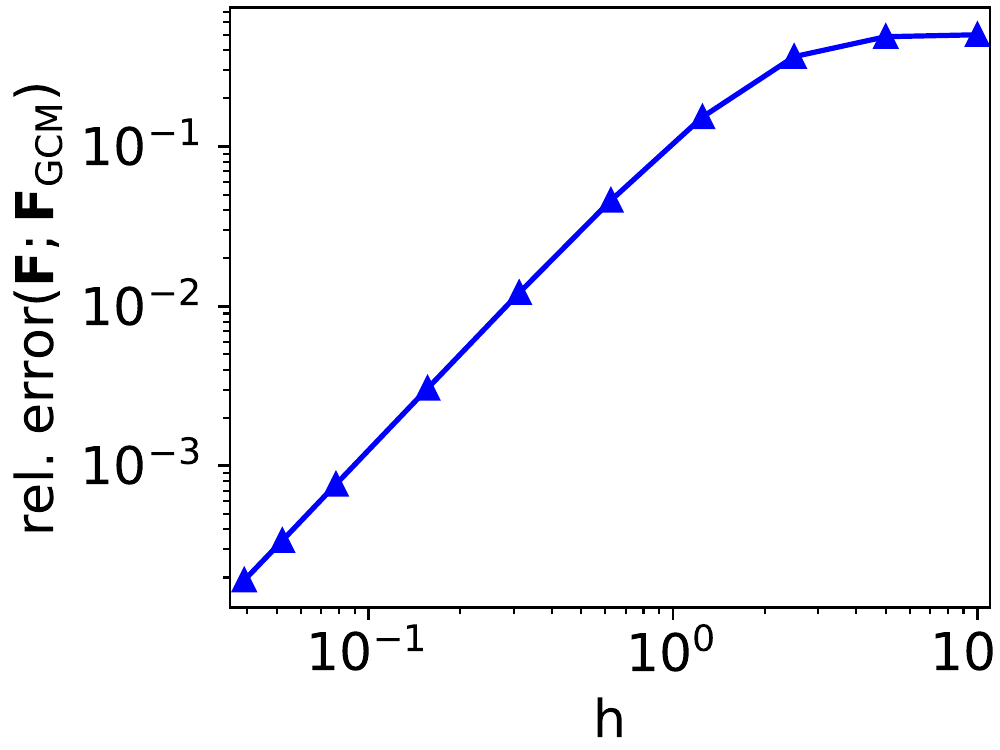}
    \caption{Relative errors of the calculated HhPF forces w.r.t.\ the pure GCM forces for varying grid spacings. For all grid spacings, the relative errors were averaged over $46$ different reduced density values in the range $0.1\le \rho^*\le 0.66$.}
    \label{fig:force_err}
\end{figure}

Having established the correspondence between explicit pair-potentials and HhPF for static properties, it is of further interest to investigate the matching dynamical properties. To gauge this, the forces must necessarily match between the two schemes. In order to thoroughly examine the extent to which pair-potential derivatives are recovered by the HhPF scheme, we turn our attention, in the rest of the present work, to only consider the GCM correspondence.  The relative differences in the calculated forces for varying grid spacings, $h$, are displayed in Figure~\ref{fig:force_err}. For each grid spacing used, the forces are averaged over all particles in each of a collection of systems of $46$ different reduced densities between $0.1\le\rho^*\le0.66$. As the differences of the total forces on all particles in the simulation box is compared, the steadily decreasing value indicates clearly that the calculated HhPF forces are fully rototranslationally invariant, as previously indicated by us.\cite{bore2020hamiltonian,ledum2020automated} The log-log slope of the relative error approaches $1.99\pm0.01$, indicating a quadratic dependence, $\mathrm{rel.\,error}(\mathbf{F};\mathbf{F}_\mathrm{GCM})\sim h^{2}$.

\begin{figure}[t]
    \centering
        \includegraphics[width=0.9\textwidth]{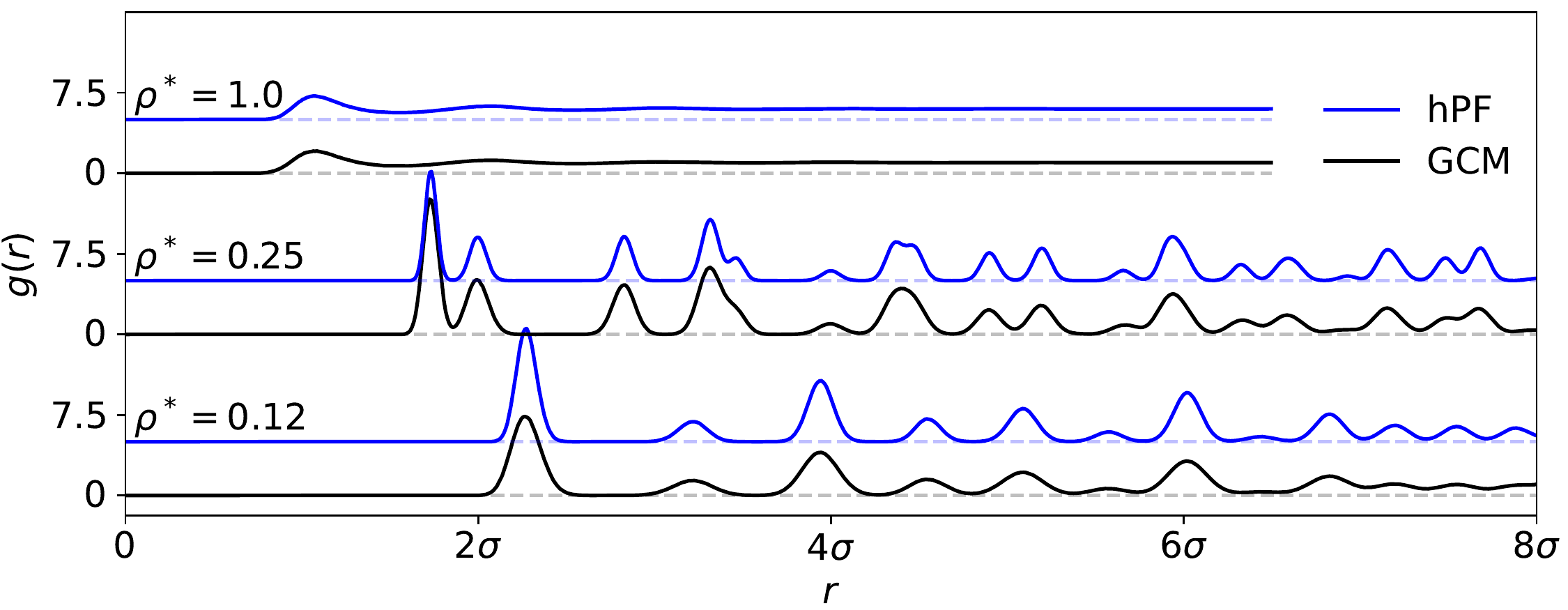}
        \caption{Radial distribution functions for GCM simulations with comparisons for corresponding HhPF at reduced densities, $\rho^*$ with reduced temperature $T^*=0.001$. Top: Liquid. Middle: Stable BCC crystal conformation. Bottom: Stable FCC crystal conformation.}
    \label{fig:gr}
\end{figure}

 With energies and forces matching between the GCM and the HhPF scheme essentially to arbitrary precision, we may now consider structural properties of the Gaussian filtered HhPF model. In Figure~\ref{fig:gr} we present radial distribution functions for select reduced system densities, representing a face-centered cubic (FCC) crystal ($\rho^*=0.12$), a body-centered cubic (BCC) crystal ($\rho^*=0.25$), and a fluid state ($\rho^*=1.0$). The former two are known to be stable solids in the GCM at the specified reduced temperature $T^*=0.001$.\cite{stillinger1976phase,prestipino2005phase} Note that the structure is excellently reproduced in the HhPF scheme, even using modest grid spacings between $h=0.352$ (for the FCC crystal) and $h=0.1771$ (for the liquid). To the best of our knowledge, this represents the first-ever hybrid particle\textendash field simulations of thermodynamically stable solids. 



\begin{figure}[t]
    \centering
        \includegraphics[width=0.40\textwidth]{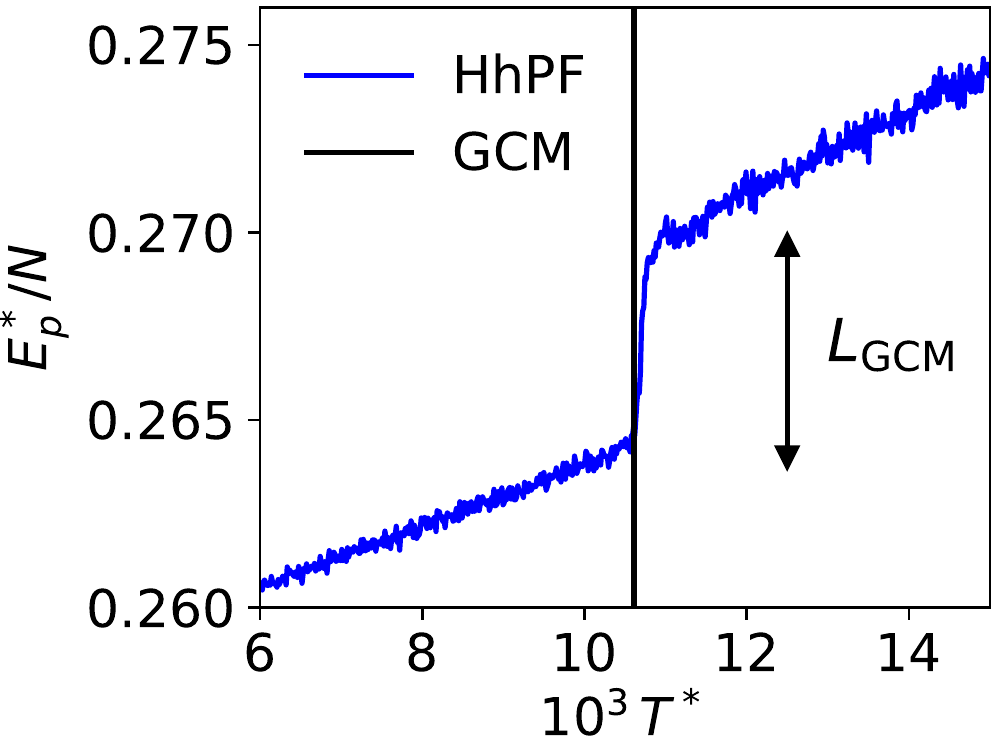}
    \caption{Reduced potential energy, $E_p^*=E_p/\varepsilon$, per particle for a slowly cooling monoatomic fluid of reduced density $\rho^*=0.25$. The red line indicates the reduced temperature where the GCM liquid\textendash solid phase transition (to BCC crystal conformation) happens. The black inset indicates the latent heat of melting of the GCM model.}
    \label{fig:phase_transition}
\end{figure}

The FCC\textendash BCC\textendash liquid transition progression for isothermal compression is typical of the GCM.\cite{stillinger1976phase,prestipino2005phase} In fact, Stillinger showed already in the 1970s that the model exhibits negative thermal expansion and a reentrant compression melting phenomenon akin to water in the $250\,\mathrm{K}\lesssim T\lesssim 273\,\mathrm{K}$ range. Because constant temperature simulations are easier to perform than constant pressure simulations in the HhPF framework, we consider in detail the melting of the BCC crystal at fixed density $\rho^*=0.25$. Shown in Figure~\ref{fig:phase_transition} is the reduced potential energy per particle of a liquid being slowly cooled. The sharp sigmoidal shape of the plot is the unmistakable fingerprint of a first-order phase transition.  The HhPF model freezes at $T^*=0.0107\pm0.0001$. This is in excellent agreement with a small GCM test system in the BCC conformation which melts at $T^*=0.0106$ upon slow heating, as reported by Prestipino et al. The reduced latent heat of melting per particle, $L/\varepsilon$, is calculated for the GCM change at $L_\mathrm{GCM}=0.00566$, which is shown inset in Figure~\ref{fig:phase_transition}, also matching the melting heat in the HhPF model.

\begin{figure}[t]
    \centering
        \includegraphics[width=0.40\textwidth]{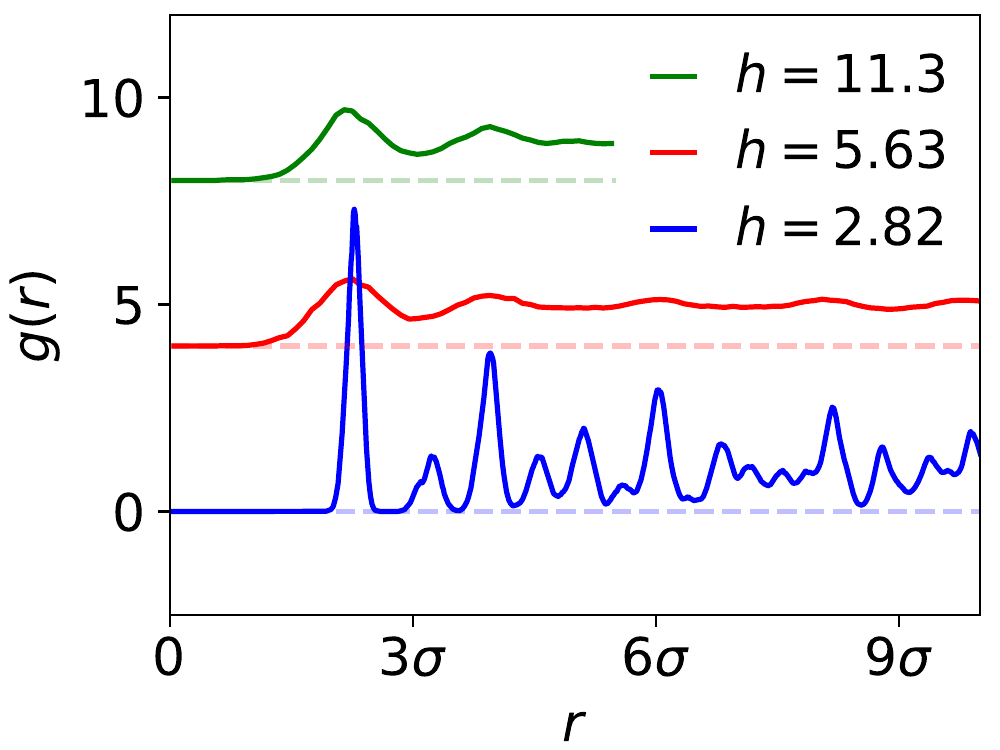}
        \includegraphics[width=0.40\textwidth]{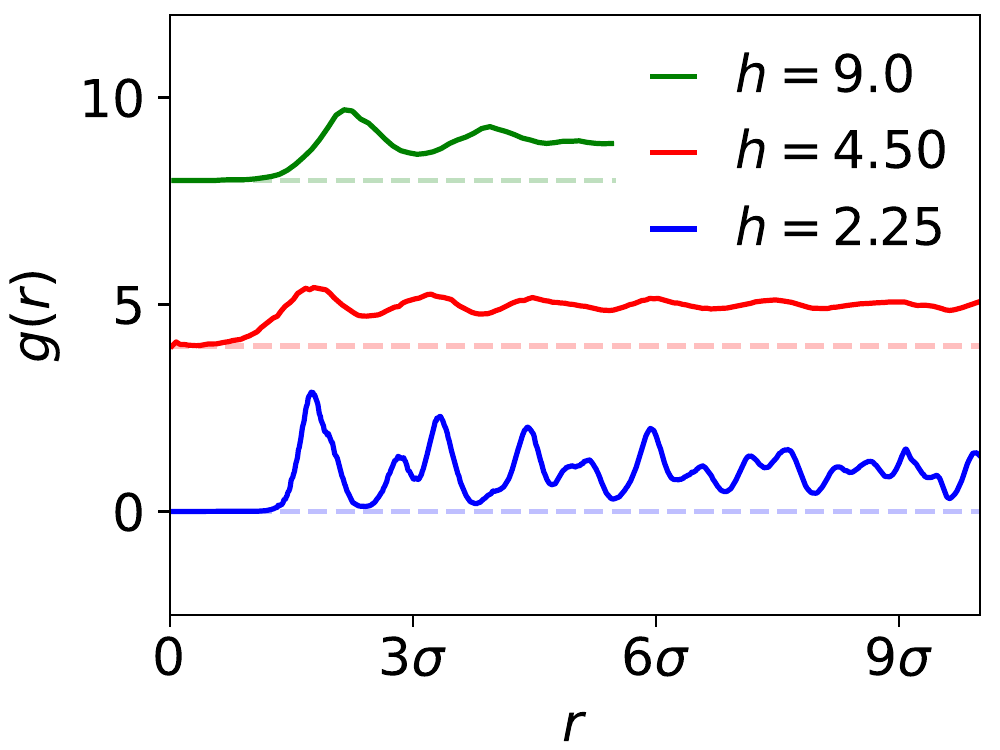}
    \caption{Radial distribution functions for HhPF of varying grid spacings, $h$. Left: Reduced density $\rho^*=0.12$ with reduced temperature $T^*=0.001$. Right: Reduced density $\rho^*=0.25$ with reduced temperature $T^*=0.0014$.}
    \label{fig:gr_grid_spacing}
\end{figure}

\paragraph{Grid dependency of the interaction strength} Traditional hPF models rely on the use of an intrinsically coarse grid to define the density fields. It is thus interesting to examine the dependence on the equivalence of the grid spacing $h$. Whereas small grid spacings yield $g(r)$-s nearly indistinguishable from the GCM, in the extreme case of a single grid point across the entire simulation box, the field is necessarily constant, and no structure is possible. Thus it is reasonable to assume a gradual softening of the field interactions as $h$ is enlarged. This effect is illustrated by Figure~\ref{fig:gr_grid_spacing}, where we note that increasing the grid spacing beyond $\sim3$ results in the crystalline conformations previously shown no longer being stable, with the system taking on an ordered fluid phase state. An additional increase in $h$ results in further loss of order, but the clear correlation hole around $g(0)$ is still present.

\paragraph{Reduced hPF units} It is clear from equation \ref{eq:intphiphi} that for any change in the HhPF filtering width $\sigma_\mathrm{hPF}\rightarrow \bar{\sigma}_\mathrm{hPF}$, a corresponding scaling of the strength of the interaction by $\sigma_\mathrm{hPF}^3/\bar{\sigma}_\mathrm{hPF}^3$ leaves the effective reduced temperature constant while scaling the reduced density by $\bar{\sigma}_\mathrm{hPF}^3/\sigma_\mathrm{hPF}$. In order to probe the dynamics of the HhPF model for varying filtering widths, it is thus sufficient to vary the reduced density $\rho^*$. 

\begin{figure}[t]
    \centering
        \includegraphics[width=0.40\textwidth]{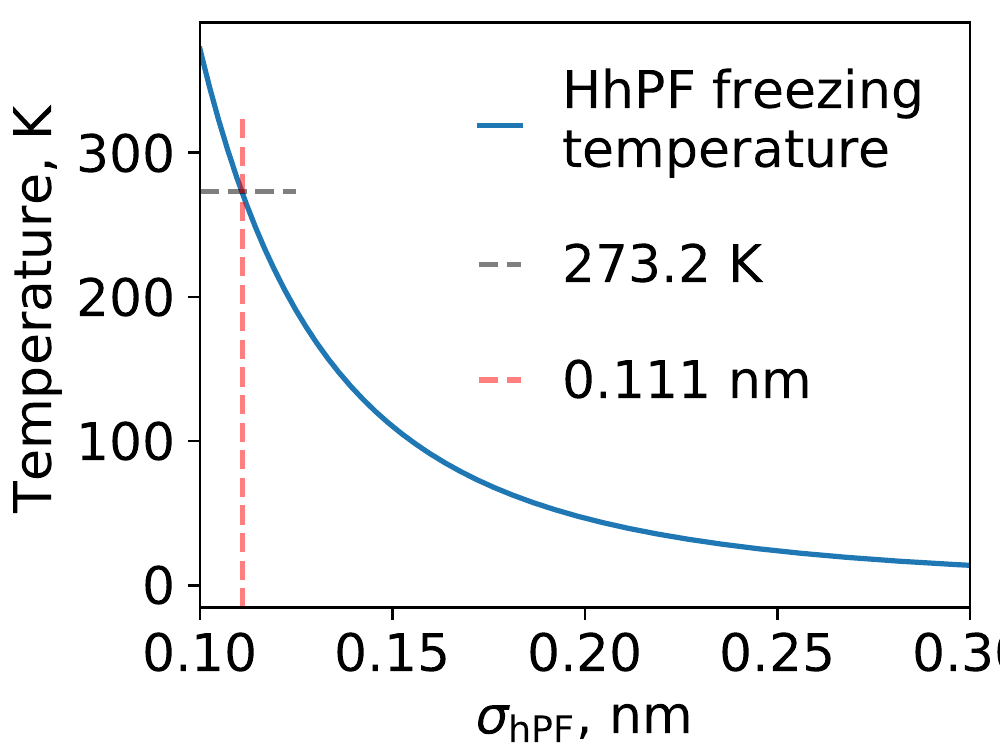}
    \caption{HhPF freezing temperature of coarse-grained water at density $\rho=10^3\,\mathrm{kg}\,\mathrm{m}^{-3}$ and $1\,\mathrm{atm.}$ pressure versus the filtering width, $\sigma_\mathrm{hPF}$.}
    \label{fig:water_freezing}
\end{figure}

In order to exemplify what typical conditions in hPF simulations correspond to in terms of GCM reduced properties, let us consider a simulation box of water at ambient conditions. In the commonly used Martini\cite{Marrink2007JCPB} four-to-one heavy atoms coarse-graining level, this water is represented by coarse-grained beads of number density $\rho=8.33\,\mathrm{nm}^{-3}$. With a filtering width around $\sigma_\mathrm{hPF}\approx 0.25\,\mathrm{nm}$, which has previously been shown to approximate well the unfiltered canonical hPF,\cite{ledum2021hylleraasmd} this corresponds to $\sigma\approx 0.5$. Using $\kappa^{-1}\approx 8RT$, which has been shown to emulate the density fluctuations of particle\textendash particle coarse-grained MD simulations,\cite{Nicolia2011JCTC} this corresponds to $\rho^*\sim1$ and $T^*\sim 0.01$. Thus the conditions illustrated in the present work are indicative of typical conditions simulated under the HhPF framework. With this filtering width, the freezing temperature of water is around $28\,\mathrm{K}$. In order to obtain a freezing temperature for the coarse-grained water of $273.2\,\mathrm{K}$, a filtering width of $\sigma_\mathrm{hPF}\approx 0.111\,\mathrm{nm}$ is necessary, see Figure~\ref{fig:water_freezing}.

\begin{figure}[t]
    \centering
        \includegraphics[width=0.30\textwidth]{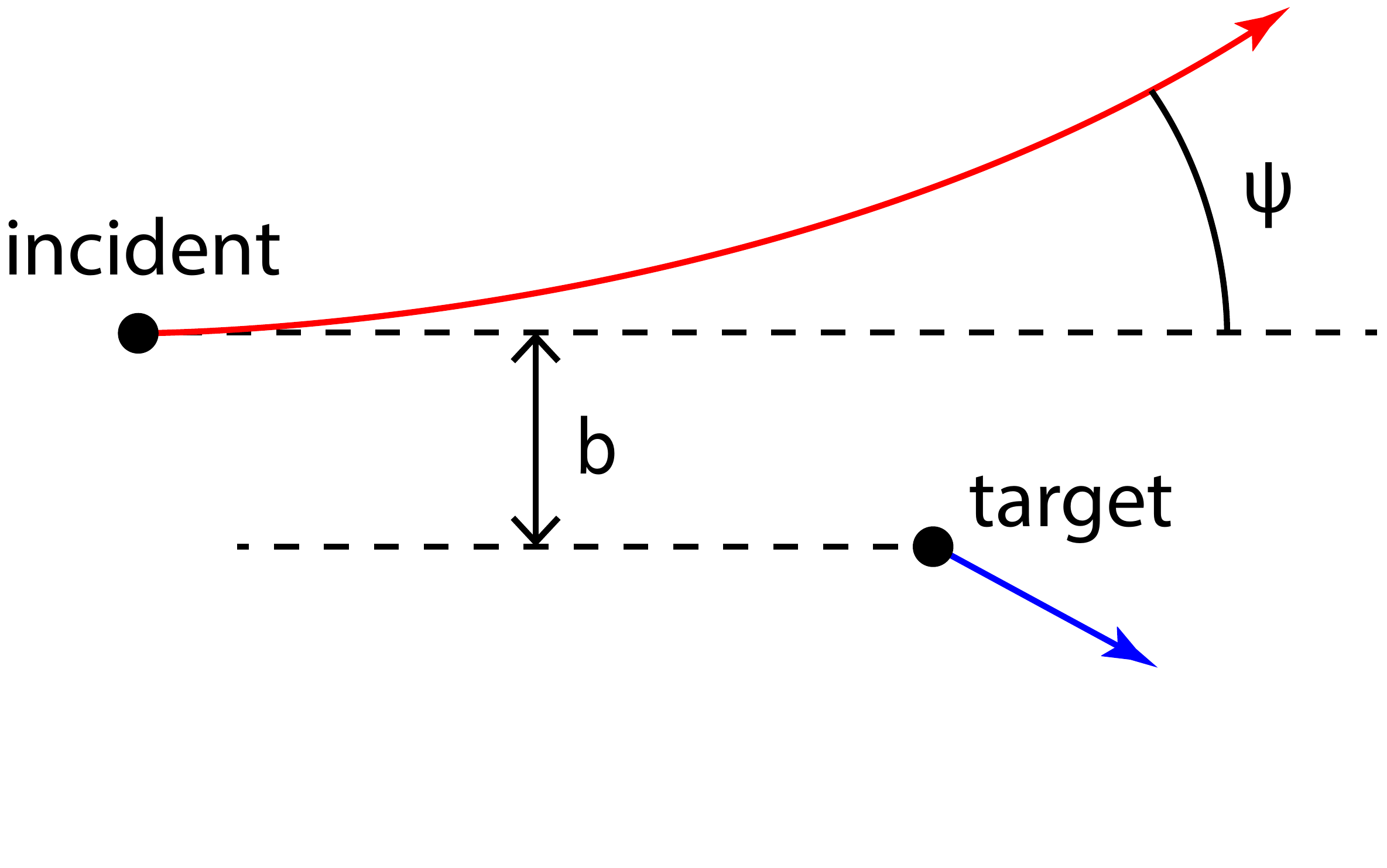}
        \includegraphics[width=0.30\textwidth]{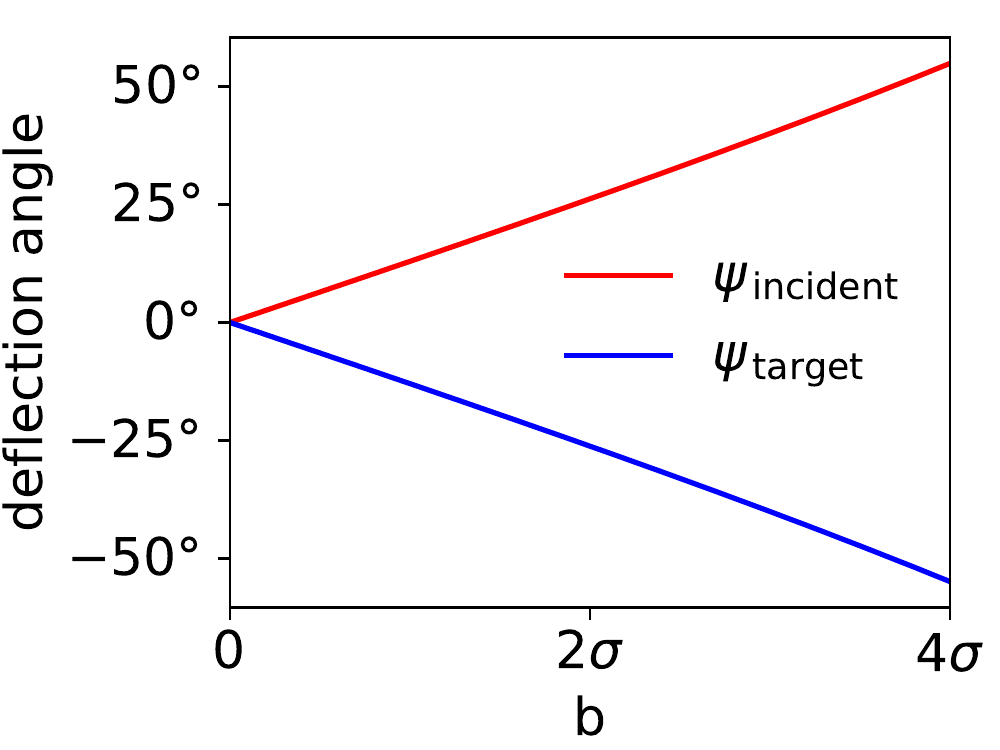}
        \includegraphics[width=0.30\textwidth]{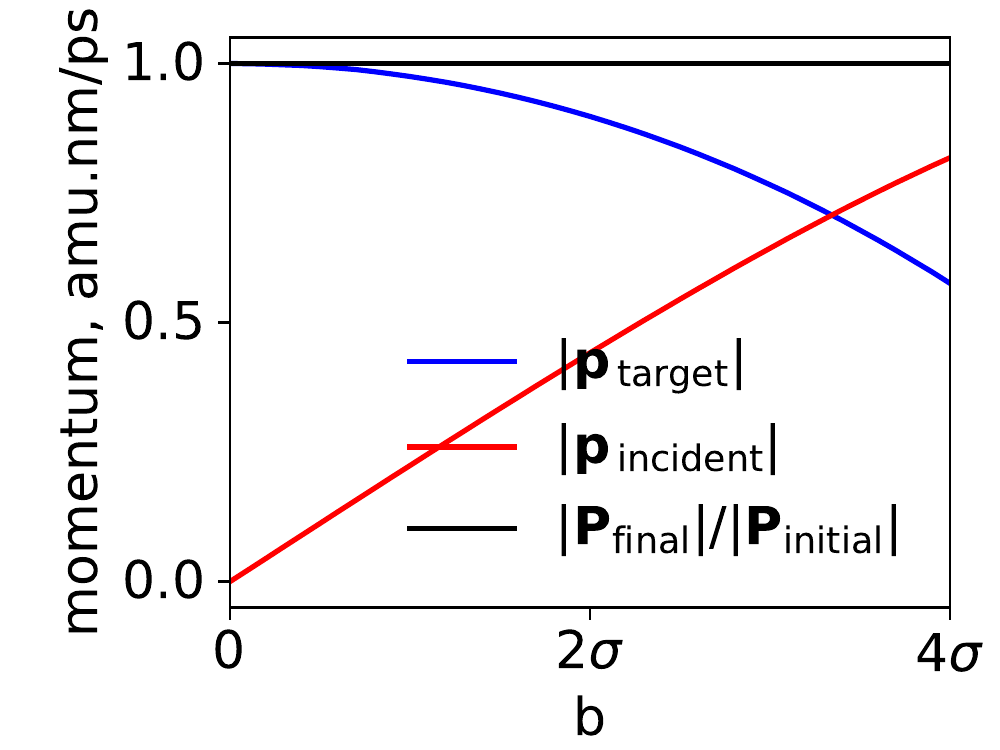}
    \caption{Deflection angle $\psi$ and momentum transfer as a function of the impact parameter $b$ in simple two-particle scattering events in the lab frame, using equal-mass particles. Low-temperature conditions are used, meaning $E_\mathrm{K}\ll \varepsilon$, i.e.\ the kinetic energy is not high enough to overcome the energy at full particle-particle overlap.
    }
    \label{fig:scattering}
\end{figure}

\paragraph{Momentum transfer} We now demonstrate further evidence that the microscopic dynamics of the HhPF is equivalent to that of the GCM, by considering the collisions of two particles compared across the two different models. First, Because HhPF is a soft-core model, we expect two different scattering regimes, depending on the kinetic energy of the system. Considering a particle with kinetic energy $E_k=p^2/2m$ colliding into a resting particle with an impact parameter $b$ (Figure~\ref{fig:scattering}).  As long as $E_k$ is smaller than the potential energy associated with full particle\textendash particle overlap, $V_\mathrm{ext}(0)$, the $b=0$ scattering event will be fully elastic, with the entirety of the incident momentum being transferred to the target particle. In the case of $E_k>V_\mathrm{ext}(0)$, the incident particle will pass \textit{through} the target particle, only imparting some fraction of the momentum depending on the $E_k/V_\mathrm{ext}(0)$ ratio. In any case, HhPF equations guarantee the correct microscopic dynamics with the rigorous conservation of the total momentum (Figure~\ref{fig:scattering}). 


\begin{figure}[t]
    \centering
        \includegraphics[width=0.4\textwidth]{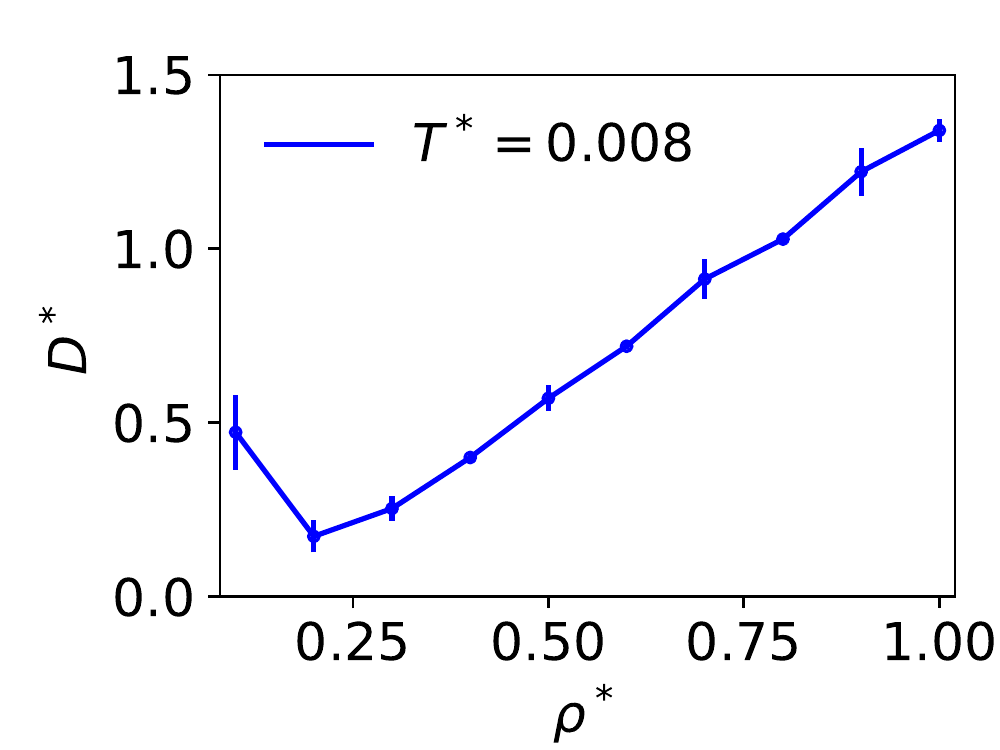}
        \includegraphics[width=0.4\textwidth]{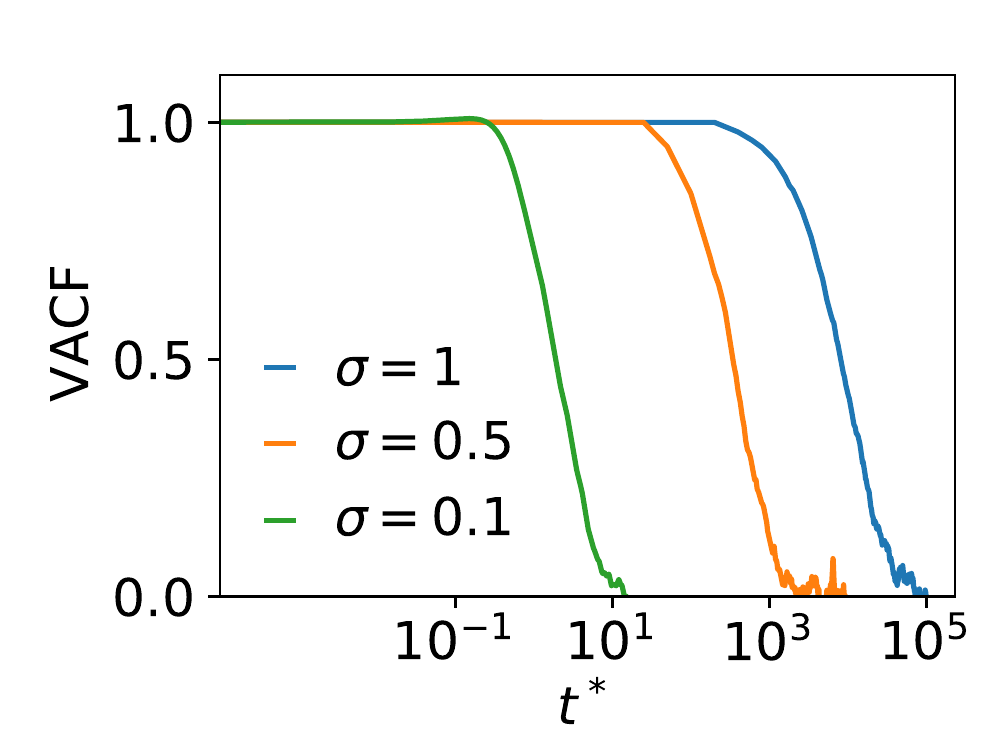}
    \caption{Left: Calculated HhPF diffusion constants, $D^*$, for varying reduced densities $\rho^*$ at reduced temperature $T^*=0.008$. Right: Normalized velocity auto-correlation functions versus reduced time, $t^*$, for varying HhPF filtering widths, $\sigma_\mathrm{hPF}$.}
    \label{fig:diffusion_constant}
\end{figure}

\paragraph{Transport properties} As a final demonstration, we consider that the non-trivial temperature dependence of the HhPF diffusion coefficient reported in Ref.\citenum{bore2020hamiltonian} associated with the scattering properties of the particles in the system is fully consistent with the behaviour of the GCM. In the low-density, low-temperature limit, the GCM exhibits a decreasing diffusion rate as the density is increased, as expected of a dilute fluid model. However, around the critical reduced density $\rho^*=\pi^{-3/2}$, an inflection point is encountered, and the trend reverses, yielding increasing self-diffusion as the reduced density increases.\cite{Mausbach2006FPE} Figure \ref{fig:diffusion_constant} (left) displays the equivalent behaviour under the HhPF method. A matching initial decrease followed by an approximately linear increase has previously been reported for the GCM by Mausbach and May.\cite{Mausbach2006FPE,sun2020self} The diffusion constant $D^*$ is expected to continue increasing upon further compression, at which point the GCM approaches a so-called "infinite density ideal-gas limit."\cite{lang2000fluid}

Also shown in Figure~\ref{fig:diffusion_constant} (right) is the normalized velocity autocorrelation function (VACF) for a set of different HhPF filtering widths. As mentioned, the state of the HhPF system is fully determined by the reduced density and temperature. However, it may be more illuminating to consider the case of constant (not reduced) density $\rho$, and varying $\sigma_\mathrm{hPF}$, as would be natural in an hPF-MD simulation setting. We note that increasing the filtering width rapidly eliminates any kind of dynamical structure, as the VACF is constant until $t>10^3t^*$. However, in the small-$\sigma_\mathrm{hPF}$ regime, clear correlations are recovered as the characteristic time of the mean free path approaches $1~t^*$.

\section{Discussion and conclusion}
The lack of excluded volume and steric effects in the hPF model can, on the one hand, be considered a strength; the smoothness of the hPF potential allowing particle overlap enables simulations to easily avoid kinetic traps which plague traditional MD simulations of bio-matter self-assembly. On the other hand, it severely limits the ability to model certain important biological soft-matter effects. For example, the phospholipid lamellar gel phase, important for describing cholesterol-rich lipid microdomains on the cell surface (so-called lipid rafts), is believed to be important for cellular signaling.\cite{simons2000lipid,lingwood2010lipid} It is therefore of great interest to thoroughly elucidate the extent to which correlations are present in hPF schemes and what determines their emergence. 
The results presented here clarify that the HhPF does not introduce any mean-field approximation. On the contrary, the model reproduces exactly the behaviour of a system of interacting particles via finite two-body potentials. Thus, HhPF models must rigorously respect all conservation laws of mechanics, prominently the conservation of the total energy and of the momentum. Moreover, colliding particles must exchange momentum to a maximum possible value dependent on the potential energy at the maximal particle-particle overlap. 

The possibility of changing the strength of particle pair-correlations in HhPF through the tunable length scale parameter $\sigma$ makes it an attractive complementary alternative to traditional CG simulation strategies which rely on pair-potentials such as the Martini model.\cite{Marrink2007JCPB,de2013improved,souza2021martini} In fact, by changing the coarse-graining parameter \textit{on the fly} it would be possible to e.g.\ quickly reach near-equilibrium self-assembled bio-matter structures and then "turn on" the correlations before the start of more accurate thermodynamic sampling.

We notice that the more general HhPF formalism fully incorporates the traditional hPF. In fact, in hPF the density functions built on coarse grids by direct CIC operation may be also defined by the convolution operation (equation~\ref{eq:phi}),  by first writing the unfiltered density as a sum of Dirac's deltas centred at the various atom positions:
\begin{equation}
    \phi(\mathbf{r}) = \sum_{i=1}^N \delta (\mathbf{r}-\mathbf{r}_i)
\end{equation}
and by defining the filter function as a triangular distribution, to be estimated on the coarse grid of choice. The same effect could also be achieved by the normal CIC assignment, with a filter $\mathcal{G}=\delta(\mathbf{x})$. Thus, as much as for HhPF, hPF schemes must respect all the conservation laws of particle-interacting  systems. In this respect, a limitation of hPF lies in the direct coupling of the numerical derivatives of the computational grid to the particle forces. This implies that changing the grid-spacing changes the range and strength of intermolecular forces. As shown in this work, the use of coarse grids for a dense system is associated with an intrinsic weakening of the interactions, systematically placing hPF in a regime of loose correlation, thus in a condition where the behaviour {\it resembles} that of a mean-field model.

In hPF, the transition from a mean field-like regime to one in which correlations arise via grid refinement would necessitate systematic re-optimization of the $\chi$ interaction parameters. Although an approach to efficiently perform such optimization has recently been suggested by us,\cite{ledum2020automated} it nevertheless remains a costly process. This issue is massively exacerbated in the case of constant pressure simulations, in which the simulation box volume is allowed to change, which, in principle, changes the canonical hPF potential every single box rescaling step. In this respect, the HhPF formalism promises a more solid theoretical basis for the determination of  the properties of systems subjected to any pressure coupling. 

In general, phase coexistence is impossible to achieve in a model with monoatomic systems with purely repulsive forces, such as the GCM, because the equation of state can be shown to be monotonic.\cite{mladek2005thermodynamic} Recently, Sevink et al.\cite{sevink2020efficient} proposed a modified hPF Hamiltonian in order to achieve this, interchanging the quadratic incompressibility term for a Cell Model or a Carnahan–Starling term. In the framework presented in this work, the futility of phase coexistence within the hPF is further made clear as the two terms in the Hamiltonian turn out to be functionally equal (up to a multiplicative constant). However, as we have demonstrated, a change of the Hamiltonian is, in fact, not necessary\textemdash a small change in the filter would be sufficient, e.g.\ using the sum
\begin{equation}
    H(\mathbf{x})\propto \varepsilon\exp\left(-\frac{\mathbf{x}^2}{\sigma^2}\right) - \exp\left(-\frac{\mathbf{x}^2}{\Sigma^2}\right),
\end{equation}
where $\Sigma>\sigma$ and $\varepsilon>1$. This would introduce a Lennard-Jones-like long-range attraction with a (finite) hard-core repulsion at close distances, opening up the possibility of a van der Waals loop enabling liquid\textendash vapour coexistence. 



Finally, in closing, we point out a possibility opened up by the work presented here. As the HhPF can be rigorously equated to a pair-interaction model, this potentially enables the employment of bottom-up coarse-graining techniques for parameter optimization. These techniques, e.g.\ force matching,\cite{izvekov2005a} were previously thought unsuited for hPF because the model was not expected to meaningfully be able to recover microscopic non-bonded interactions from pair-potentials. Currently, the only proposed automatic algorithm for finding $\tilde\chi_{k\ell}$-parameters in the literature is a derivative-free black box optimization based on Bayesian optimization.\cite{ledum2020automated} The ability to use more conventional bottom-up strategies has the potential to reduce the computational cost of hPF parameter optimization by several orders of magnitude.




\section{Data availability statement}
The full code used to produce all the results contained in this work is free and openly available at the GitHub repository of the HylleraasMD code, \url{https://github.com/Cascella-Group-UiO/HyMD}.

\section{Acknowledgements}
This work was supported by the Research Council of Norway through the Centre of Excellence \textit{Hylleraas Centre for Quantum Molecular Sciences}  (grant number 262695), by the Norwegian Supercomputing Program (NOTUR) (grant number NN4654K), and by the Deutsche Forschungsgemeinschaft (DFG) within the project B5 of the TRR-146 (project number 233630050).

\bibliography{ref}

\end{document}